\documentclass[9pt,twocolumn,twoside]{pnas-new}

\templatetype{pnasresearcharticle} % Choose template
\begin{document}

\title{Simultaneously Infer Cell Pseudotime,Velocity Field and Gene Interaction from Multi-Branch scRNA-seq Data with scPN}

% Use letters for affiliations, numbers to show equal authorship (if applicable) and to indicate the corresponding author
\author[a]{Zhen Zhou}
\author[a,1]{Jiachen Li}
\author[b]{Hongyi Xin}
\author[a]{Xiaoyong Pan}
\author[a,1]{Hong-Bin Shen}

\affil[a]{Institute of Image Processing and Pattern Recognition, Shanghai Jiao Tong University, and Key Laboratory of System Control and Information Processing, Ministry of Education of China, Shanghai, 200240, China}
\affil[b]{Global Institute of Future Technology, Shanghai Jiao Tong University, Shanghai, 200240, China}

% Please give the surname of the lead author for the running footer
\leadauthor{Zhou}

% Please add a significance statement to explain the relevance of your work
\significancestatement{
% Combining temporal information with the gene interaction matrix enables dynamic analysis of gene regulation and cell fate, offering a more comprehensive and profound understanding of cellular behavior. It is challenging to simultaneously infer cell pseudotime and gene interaction networks, especially in multi-branch differentiation scenarios. 
% Currently, common methods for calculating cellular dynamics first compute single-cell velocity and then infer pseudotime based on this velocity. 
% However, this sequential approach may lead to inconsistencies between the pseudotime and the velocity. 
% Our method, scPN, addresses this issue by simultaneously optimizing both pseudotime and velocity, ensuring consistency between the two. 
% The single-cell Piecewise Network (scPN) uses a piecewise Ordinary Differential Equation (ODE) model to iteratively extract temporal patterns and inter-gene relationships. scPN uniquely reconstructs both pseudotime and gene interaction networks, enhancing accuracy in predicting dynamic cellular changes and uncovering distinct branching paths. Experimental results demonstrate scPN's superior performance in reconstructing cellular dynamics and identifying key transcription factors, offering a significant advancement in understanding cellular development and gene regulation
The Single-cell Piecewise Network (scPN) addresses the challenge of simultaneously inferring pseudotime and gene interaction networks in multi-branch differentiation. Unlike traditional methods that first compute single-cell velocity and then infer pseudotime, leading to potential inconsistencies, scPN optimizes both pseudotime and velocity concurrently. By using a piecewise Ordinary Differential Equation (ODE) model, scPN iteratively extracts temporal patterns and inter-gene relationships, ensuring consistency between pseudotime and velocity. This method reconstructs both pseudotime and gene interaction networks more accurately, uncovering branching paths and predicting dynamic cellular changes. Experimental results highlight scPN's superior ability to reconstruct cellular dynamics and identify key transcription factors, advancing the understanding of cellular development and gene regulation.}

% Please include corresponding author, author contribution and author declaration information
\authorcontributions{Author contributions: H.S. and J.L. designed research; Z.Z. performed research; H.X. and X.P. contributed new reagents/analytic tools; Z.Z. analyzed data; and Z.Z., J.L. and H.S. wrote the paper.}
\authordeclaration{The authors declare no competing interests.}
\correspondingauthor{\textsuperscript{1}To whom correspondence should be addressed. E-mail: lijc0804@sjtu.edu.cn (J.L.) and hbshen@sjtu.edu.cn (H.S.)}

% At least three keywords are required at submission. Please provide three to five keywords, separated by the pipe symbol.
\keywords{single-cell analysis $|$ Pseudotime inference $|$ Piecewise ODE $|$ Gene interaction network }

\begin{abstract}
Modeling cellular dynamics from single-cell RNA sequencing (scRNA-seq) data is critical for understanding cell development and underlying gene regulatory relationships. Many current methods rely on single-cell velocity to obtain pseudotime, which can lead to inconsistencies between pseudotime and velocity. It is challenging to simultaneously infer cell pseudotime and gene interaction networks, especially in multi-branch differentiation scenarios. We present single-cell Piecewise Network (scPN), a novel high-dimensional dynamical modeling approach that iteratively extracts temporal patterns and inter-gene relationships from scRNA-seq data. To tackle multi-branch differentiation challenges, scPN models gene regulatory dynamics using piecewise gene-gene interaction networks, offering an interpretable framework for deciphering complex gene regulation patterns over time. Results on synthetic data and multiple scRNA-seq datasets demonstrate the superior performance of scPN in reconstructing cellular dynamics and identifying key transcription factors involved in development compared to existing methods. 
To the best of our knowledge, scPN is the first attempt at modeling that can recover pseudotime, velocity fields, and gene interactions all at once on multi-branch datasets.

\end{abstract}

\dates{This manuscript was compiled on \today}
\doi{\url{www.pnas.org/cgi/doi/10.1073/pnas.XXXXXXXXXX}}

\maketitle
\thispagestyle{firststyle}
\ifthenelse{\boolean{shortarticle}}{\ifthenelse{\boolean{singlecolumn}}{\abscontentformatted}{\abscontent}}{}

\firstpage[5]{3}
% Use \firstpage to indicate which paragraph and line will start the second page and subsequent formatting. In this example, there are a total of 11 paragraphs on the first page, counting the first level heading as a paragraph. The value {12} represents the number of the paragraph starting the second page. If a paragraph runs over onto the second page, include a bracket with the paragraph line number starting the second page, followed by the paragraph number in curly brackets, e.g. "\firstpage[4]{11}".

% If your first paragraph (i.e. with the \dropcap) contains a list environment (quote, quotation, theorem, definition, enumerate, itemize...), the line after the list may have some extra indentation. If this is the case, add \parshape=0 to the end of the list environment.
\dropcap{T}he emergence of scRNA-seq has significantly enhanced our ability to investigate cellular dynamics. While providing insights into cell subpopulations and uncovering factors influencing cellular state shifts, scRNA-seq captures only static snapshots of cell populations and lacks explicit demonstration of dynamic transitions between cellular states. In response to this limitation, trajectory inference algorithms like slingshot \cite{street2018slingshot}, monocle \cite{trapnell2017monocle}, Palantir \cite{setty2019characterization} have been developed to construct potential branching trajectories based on transcriptomic profile similarities. Meanwhile, RNA velocity \cite{la2018rna} theory was proposed to predict cell fate by modeling the dynamics between the unspliced  and spliced mRNAs with differential equations. More recently, considerable effort has been dedicated to refining these methods in order to enhance parameter inference \cite{bergen2020generalizing}, mitigate the impact of noisy measurements associated with individual features \cite{lange2022cellrank}, construct dynamic models from transcriptional regulation \cite{li2024tfvelo}, capture the uncertainty in splicing process \cite{gayoso2024deep}, and incorporate multiview learning \cite{weiler2024cellrank}. While these approaches have brought valuable biological insights into cell differentiation and disease progression, most existing models assume uniform kinetics across all cells in an scRNA-seq experiment \cite{li2024relay}. This assumption may lead to poor predictive performance when cell subpopulations exhibit dissimilar RNA velocity kinetics, a scenario often encountered in complex biological conditions.

The rapid advancement of single-cell genomics has unveiled extensive cellular heterogeneity in processes such as organismal development, regeneration, and disease states. This diversity is primarily orchestrated by the synergistic actions of transcription factors (TFs) \cite{holland2020robustness, bartosovic2021single}, which modulate the activation or repression of expressed genes by binding to regulatory regions of the genome. As TFs are subjected to their own regulatory mechanisms, dynamic changes in cellular phenotype can be achieved by modifying TF activity in response to intrinsic or extrinsic signals. Thus, a comprehensive understanding of both static and dynamic cellular heterogeneity can be gleaned from the study of differential TF activity over time and across phenotypic space.

Numerous methods have been devised to measure the regulatory activity from scRNA-seq data, often by inferring gene regulatory networks (GRNs) associating TFs with their regulated genes (target genes). Whether these associations are deduced from gene expression data \cite{aibar2017scenic}, epigenetic data \cite{dong2022single}, or a combination of both \cite{jansen2019building}, many such approaches typically consider static, predefined cell clusters as the endpoint of differentiation. For instance, CellOracle \cite{kamimoto2023dissecting} presents an approach for network inference combined with simulated gene perturbation to unravel cell identity. GRouNdGAN \cite{zinati2024groundgan} is a GRN-guided, reference-based causal implicit generative model. It is designed to simulate single-cell RNA sequencing data, conduct in silico perturbation experiments, and benchmark GRN inference methods. SCENIC+ \cite{bravo2023scenic+} is a tool designed to infer enhancer-driven GRNs by integrating scRNA-seq and single-cell chromatin accessibility sequencing (scATAC-seq) data.

Recent studies have explored the modeling of GRNs by incorporating temporal patterns. Dictys \cite{wang2023dictys} employs multiomic single-cell assays to investigate chromatin accessibility and gene expression. It integrates context-specific transcription factor footprints and utilizes a stochastic process network alongside an efficient probabilistic model to analyze single-cell RNA-sequencing read counts. Symphony defines cell-type-specific programs with a Bayesian approach that models individual cells as mixtures governed by unique GRNs \cite{bachireddy2021mapping}. scKINETICS \cite{burdziak2023sckinetics} simultaneous learns the per-cell transcriptional velocities and a governing gene regulatory network, which employs an expectation–maximization approach to infer the impact of each regulator on its target genes. However, scKINETICS fails to account for the fact that different cell categories may possess distinct connectivity matrices, overlooking the potential variability in regulatory interactions among cell types.

Inspired by the idea that once a gene-gene interaction network is constructed, each cell can be seen as a snapshot of that network, we recognize that the challenge of pseudotime and gene interaction inference fundamentally involves reconstructing timestamps and connection matrices in dynamic networks. Recent studies have explored predicting network dynamics without complete observation of network attributes \cite{prasse2022predicting, gao2022autonomous}. In the gene-gene interaction network, scRNA data provides the gene expression of cells without temporal order, which could be regarded as the value of each node at unknown sampling time point. Our goal is to reconstruct the time series for each node and the interactions between them.

Due to the current single-cell velocity methods that first compute the velocity field and then derive the temporal ordering, inconsistencies arise between the pseudotime and the velocity field. Additionally, methods based on Gene Regulatory Networks (GRNs) struggle to effectively handle multi-branch datasets. To address these issues, we propose a novel approach named scPN.The scPN reconstructs the velocity field and pseudotime as well as the gene-gene interaction network by introducing a piecewise Ordinary Differential Equation (ODE) model. By optimizing the gene-gene connectivity matrices and the pseudotime of cells iteratively, scPN seeks to enhance the accuracy of predicting dynamic cellular changes. Using the piecewise network, scPN could addresses the challenges posed by branched single-cell trajectories, where cells may diverge into different lineages. This capability is crucial for understanding the complexity of cell differentiation and development, as it helps uncover distinct branching paths within single-cell trajectories and transitions in cell fate. Results on synthetic and scRNA-seq datasets show that the proposed approach could accurately infer the cellular state transitions, and detect key gene regulatory relations. To the best of our knowledge, scPN is the only method that can simultaneously infer pseudotime and the gene interaction matrix on multi-branch datasets.To the best of our knowledge, scPN is the only method that can simultaneously infer pseudotime and the gene interaction matrix on multi-branch datasets.

\section*{Methods}

\subsection*{Modeling cellular dynamics with piecewise networks}
In our study, we deal with a real single-cell gene expression matrix of size \(T \times N\), where \(T\) represents the number of cells and \(N\) denotes the number of genes. We adopt a piecewise linear and time-homogeneous system of differential equations. One function describes the rate of change of an individual gene \(i\), as expressed in the provided equation. Each cell possesses \(N\) genes, and the interactions among these \(N\) genes are captured by an \(N \times N\) matrix denoted as \(A\). The coefficient \(a_{ij}\) represents the impact of gene \(j\) on gene \(i\), while \(\frac{dx_i(t)}{dt}\) signifies the rate of change in the expression of gene \(i\).

Due to the unavailability of real-time labels in experiments, we assign time labels to the \(T\) cells by sorting them, resulting in a sorted matrix. In this sorted matrix, we enforce the condition that the expression levels of genes over time are continuous, implying that \(x(t)\) is a continuous function. In the case of a piecewise linear system of differential equations, each region corresponds to the linear system described above. We assume that throughout the entire sorted time range, the expression levels of genes in adjacent regions should be continuous, ensuring a smooth transition between them.

The structure of a graph is represented by the \(N \times N\) weighted adjacency matrix \(A\) whose elements are denoted by \(a_{ij}\). Here, we consider a fixed weighted adjacency matrix \(A\) with the underlying graph remaining constant over time.

We denote the nodal state of node \(i\) at time \(t\) by \(x_i(t)\) and the nodal state vector by \(x(t) = \left(x_1(t), \ldots, x_N(t)\right)^T\). Considering a general class of dynamical models on networks, describing the evolution of nodal state \(x_i(t)\) of any node \(i\) as
\begin{equation}
\frac{dx_i(t)}{d t} = f_i\left(x_i(t)\right) + \sum_{j=1}^N a_{ij} g\left(x_i(t), x_j(t)\right).
\end{equation}

Additionally, given the observed and disrupted node states \(x_i\), our objective is to reconstruct the dynamics of the network, including the timestamps \(t\) and the connectivity matrix \(A\).

The function \(f_i\left(x_i(t)\right)\) characterizes the self-dynamics of node \(i\). The sum in equation [1]  signifies interactions of node \(i\) with its neighbors. The interaction between two nodes \(i\) and \(j\) depends on the adjacency matrix \(A\) and the interaction function \(g\left(x_i(t), x_j(t)\right)\). A broad spectrum of models follows from equation [1]  by specifying the self-dynamics function \(f_i\) and the interaction function \(g\). For example, by setting \(f_i\left(x_i(t)\right) = 0\) and \(g\left(x_i(t), x_j(t)\right) = x_j(t)\), the equation [1] reduces to linear dynamics \(\frac{d x(t)}{d t} = A x(t)\). 

To deal with the temporal ordering of multi-branch datasets, our scPN primarily focuses on the piecewise linear model, shown in Figure 1(b), where piecewise linear models are employed for different regions of \(x\), expressed as:
\begin{equation}
\frac{d x_i(t)}{d t} = \sum_{k=1}^m \sum_{j=1}^N a_{ij}^{(k)} x_j(t) \cdot \boldsymbol{1}_ {\{x_t \in R_k\}}
\end{equation}
Here:
$\boldsymbol{1}_ {\{x_t \in R_k\}}$ is the indicator function, indicating 1 when $x_t$ is in region $R_k$ and 0 otherwise. And $x_t$  represents the all $x_i(t)$ with time stamp $t$.
\(\frac{d x_i(t)}{d t}\) represents the rate of change of node \(i\) with respect to time \(t\). The model consists of \(m\) different regions \(R_1, R_2, \ldots, R_m\), each characterized by a distinct linear equation. In each region, the contribution of neighboring nodes \(x_j(t)\) is determined by the corresponding coefficients \(a_{ij}^{(1)}, a_{ij}^{(2)}, \ldots, a_{ij}^{(m)}\). This piecewise formulation allows for different linear dynamics depending on the region in which \(x_t\) resides.
Our motivation for creating this model stems from the presence of bifurcation phenomena observed in single-cell data, particularly in the context of hippocampal dentate gyrus data. The challenge arises from the branching behavior exhibited by individual cells, leading to distinct trajectories in the data.

To deal with the heterogeneity of gene regulation relations enrolled in the development process , we introduce a piecewise model scPN where different connectivity matrix coefficients are applied to different cell clusters. This allows us to capture the diversity in gene network dynamics among various cell fates. The model accommodates the variability in the data by adapting the network structure based on the characteristics of the individual clusters, providing a more nuanced representation of the underlying dynamics in the presence of branching phenomena.

\subsection*{The framework of scPN}
In the entire workflow (Figure 1), scPN begins by standardizing the data, performing gene selection, and taking logarithms. Due to numerous dropouts in the single-cell expression matrix, we employ the AutoClass framework for imputation (Figure 1(b)). For scRNA-seq datasets which describe cell development with multiple branches, it is necessary to construct piecewise networks for modeling the various gene interaction relations across different cell types.

\begin{figure*}[h]
    \centering
    \includegraphics[width=\textwidth]{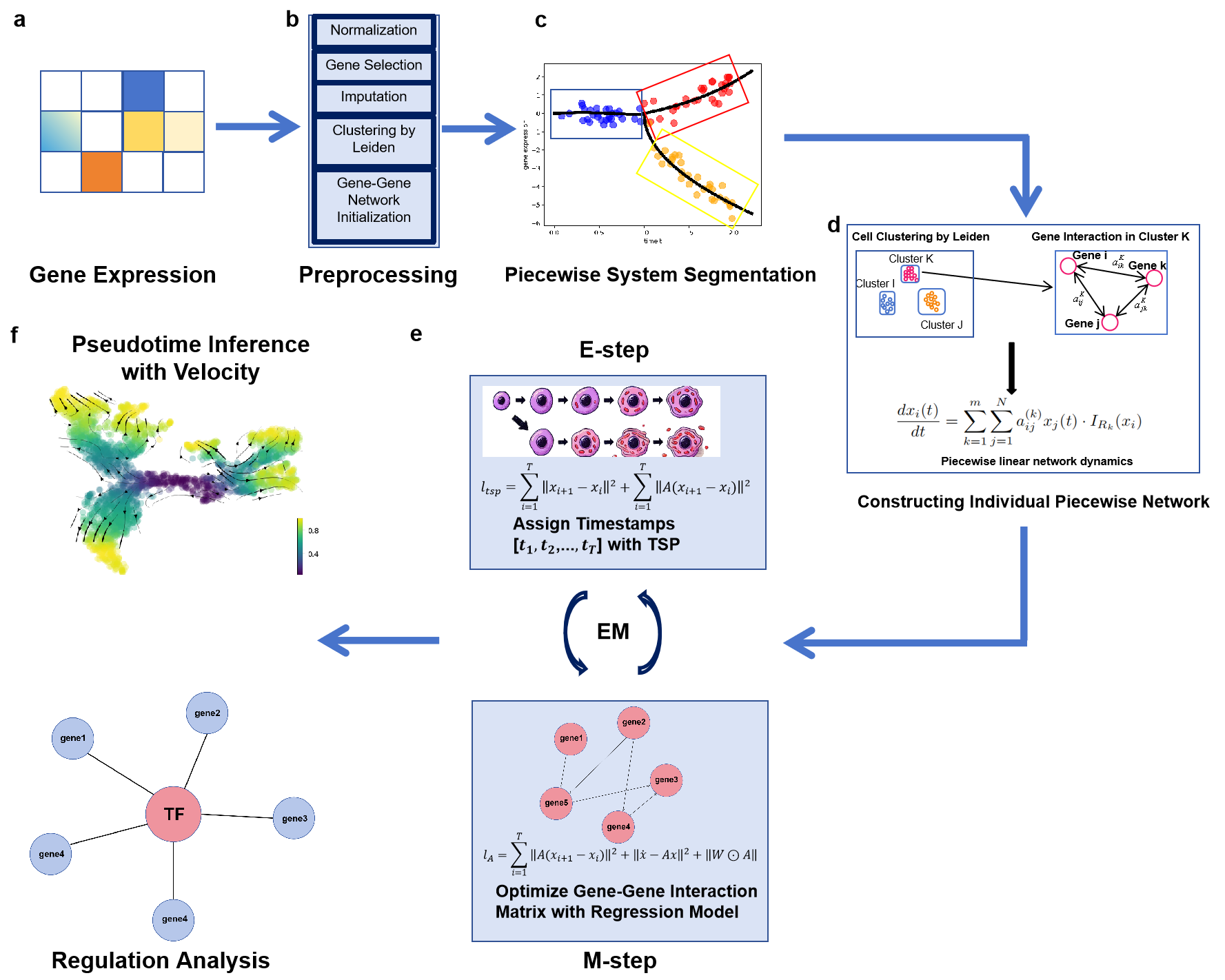}
    \caption{
\textbf{The framework of scPN, which can simultaneously obtain the temporal dynamics and the gene-gene interaction matrix.} (a) Raw dataset of gene expression. The single-cell gene expression matrix is typically of size \( T \times N \), where \( T \) represents the number of cells and \( N \) denotes the number of genes. Generally, this gene expression matrix is sparse.
(b) The preprocessing procedure of scPN. It includes normalization, gene selection, imputation, clustering, followed by piecewise linear network modeling, and initialization of the gene-gene interaction matrix with prior knowledge. (c$\sim$d) 
Constructing individual piecewise network after clustering. scPN clusters all cells using the Leiden algorithm and constructs distinct piecewise networks for each cluster, enabling the inference of gene interaction matrices across different time intervals. (e) scPN algorithm. scPN is an iterative algorithm similar to the EM algorithm. It derives pseudotime dynamics from the known gene-gene interaction matrix $A$ with TSP and infers the interaction matrix from the known temporal dynamics with regression.
(f) The output of scPN. The output results of scPN includes single-cell pseudotime, velocity field and the gene-gene weight matrix.Moreover, we provide an analysis of the gene-gene interaction matrix.
}
    \label{fig:b}
\end{figure*}

The scPN employs an Expectation-Maximization (EM) algorithm (Figure 1(e)) that iteratively update time labels by solving Traveling Salesman Problem (TSP) \cite{gavish1978travelling} and the gene-gene matrix. Prior knowledge of gene interaction relationships is optional which can enhance the interpretability of the learned gene-gene matrix. After multiple iterations, scPN can provide the inferred pseudotime for each cell and the gene interaction matrix of the dynamic network, which can detect key TFs involved in development.Finally, scPN derives the single-cell velocity from the left-hand side of Equation [2], \(\frac{d x_i(t)}{d t}\). This enables the visualization of the velocity field on the pseudotime plot
 (Figure 1(f)).

\subsection*{Preprocessing of scRNA-seq data}
The scRNA-seq data is preprocessed using the strategy widely adopted in previous studies \cite{setty2019characterization, bergen2020generalizing}. After filtering out genes with fewer than 20 expressed counts, the top 2,000 highly variable genes are selected. The gene expression counts are then log-transformed, and a nearest-neighbor graph is constructed using 30 neighbors.

To model the cell developmental process with a dynamic network,  the gene expression levels are supposed to be smooth functions of time $t$.  Because of the high noise and sparsity in scRNA-seq data,  imputation for the gene expression matrix \cite{patruno2021review} is necessary. For this purpose, we employed AutoClass \cite{li2022universal}, which leverages a unified neural network architecture combining an autoencoder and a classifier to improve imputation in single-cell data. This approach enhances noise filtering and signal retention, accommodating non-linear gene relationships, and thereby producing a more accurate and comprehensive dataset for analyzing temporal dynamics.

After imputation, for the multi-branch datasets, scPN firstly uses the Leiden algorithm to get cell clusters. Then, we model each cell type in segments, completing the piecewise linear model construction.

Finally, we need to initialize the gene interaction matrix. To enhance the interpretability of our computed gene connectivity matrix, we incorporated prior knowledge regarding transcription factors. Specifically, we utilized the ChEA \cite{keenan2019chea3} TF-target database to identify potential regulatory positions, which we denoted as 1, resulting in a 0-1 mask matrix \( W \) of size \( 2000 \times 2000 \). To ensure that this mask matrix influences our loss function, we introduced a prior loss involving the element-wise (Hadamard) product of \( W \) and the norm of \( A \):
\begin{equation}
l_{prior}= \| W \odot A \| 
\end{equation}
This approach ensures that the final connectivity matrix \( A \) only has values where \( W_{ij} \) is 1. Consequently, our algorithm no longer has an analytical solution and requires the use of gradient descent to compute \( A \). In the following sections, we will apply this method to three datasets.

\subsection*{Optimization of scPN model}
After preprocessing, scPN assumes the expression levels of genes ($x$) and their time derivative ($\frac{dx}{dt}$) are smooth functions with respect to time $t$. If the order of cells is correctly sorted,  the sum $\sum_{i=1}^{T-1} \operatorname{dist}[t_i, t_{i+1}]$ will be minimized. We can define the distance between two cells (indexed as $t_i$ and $t_j$) as follows:

\begin{align}
\operatorname{dist}[i, j] & = \left\|x\left(t_i\right) - x\left(t_j\right)\right\| + \left\|\dot{x}\left(t_i\right) - \dot{x}\left(t_j\right)\right\|  \notag\\
& = \left\|x\left(t_i\right) - x\left(t_j\right)\right\| + \left\|A x\left(t_i\right) - A x\left(t_j\right)\right\|
\end{align}

We define the sum of the above distances $\sum_{i=1}^{T-1} \operatorname{dist}[t_i, t_{i+1}]$ as the TSP loss, which addresses the continuity of $x(t)$ with respect to $t$. It ensures that the state variable $x(t)$ remains continuous over time. 
\begin{equation}
l_{TSP} = \sum_{i=1}^{T-1}\|x_{i+1}-x_i\|^2 +\sum_{i=1}^{T-1}\|A(x_{i+1}-x_i)\|^2
\end{equation}
Here,  $\sum_{i=1}^{T-1}\|A(x_{i+1}-x_i)\|^2$ emphasizes the continuity of the gradient $\dot{x}$. It ensures that the gradient remains continuous across different time steps. 

Next, we define the Regression loss as: 
\begin{equation}
    l_{regre}= \|\dot{x} - A x\|^2,
\end{equation}
which focuses on the recovery of the dynamic system. It aims to minimize the difference between the actual dynamic system represented by $\dot{x}$ and the system reconstructed using $A$ and $x$.

In conclusion, the overall loss can be denoted as:

\begin{equation}
\begin{aligned}
l &= l_{TSP} + l_{regre} + l_{prior} \\
&= \sum_{i=1}^{T-1}\|x_{i+1}-x_i\|^2 + \sum_{i=1}^{T-1}\|A(x_{i+1}-x_i)\|^2 \\ &+ \left\|\dot{x} - A x\right\|^2 + \left\| W \odot A \right\|
\end{aligned}
\end{equation}

To optimize the overall loss function, scPN adopts an EM framework. The algorithm iteratively performs two steps, each designed to reduce the overall loss, eventually converging to a local minimum. In the E step, because only  TSP loss is dependent on the pseudotime,  $t$ is updated by minimizing $l_{TSP}$. The Two-Opt algorithm \cite{wu2006revised} is employed for solving this TSP, which iteratively improves a tour by reversing the order of a chosen subset of edges to eliminate crossings and reduce the TSP Loss. In the M step,  the gene interaction matrix $A$ is optimized according to the terms related to $A$ in the loss function, which can be written as 
\begin{equation}
l_A=\sum_{i=1}^{T-1}\left\|A\left(x_{i+1}-x_i\right)\right\|^2 + \|\dot{x} - A x\|^2 + \left\| W \odot A \right\|
\end{equation}

\( A \) could be optimized numerically or analytically, depended on whether $l_{prior}$ exists. When the prior knowledge about gene-interaction is provided, scPN updates \( A \) by gradient descent of $l_A$. When there is no prior knowledge about gene interactions, which means that $l_{prior}$ is not included in the loss function, we can derive the analytical solution of $A$ by setting $\frac{\partial l_{A}}{\partial A} = 0 $:

\begin{equation}
\hat{A} = \left(\sum_{i=1}^{T-1}\left(x_{i+1}-x_i\right)^{\top}\left(x_{i+1}-x_i\right) + x^{\top} x\right)^{-1} \cdot x^{\top} \cdot \dot{x}
\end{equation}
\section*{Results}
\subsection*{Dynamics inference on synthetic networks with scPN}
In this section, we present the results of simulation experiments, encompassing the responses of four nonlinear dynamic systems and one piecewise linear system. We depict the temporal evolution of various nodes without the knowledge of connection matrices $A$ and time $t$.

For the four nonlinear dynamic systems, we first randomly generate and initialize  a connection matrix \( A \) of size \( 100 \times 100 \), where each \( A_{ij} \) denotes the connection between the \( i \)th and \( j \)th nodes in the dynamic network. Subsequently, we randomly generate an initial value \( \mathbf{x}_0 \), a 100-dimensional vector. Utilizing the equations outlined in Table 1, we can model the evolving states of 100 nodes over a time period \( T \). Afterward, we permute the \( T \) sequences of 100-dimensional vectors, resulting in a \( 100 \times T \) matrix representing the shuffled data.

In this work, we explore various models of dynamics on networks, as detailed in Table 1.

\begin{table}[h]
    \centering
    \caption{Multiple Types of Nonlinear Dynamic Networks}
    \begin{tabular}{l l}
        \toprule
        \textbf{Network}& \textbf{Dynamical Equation}\\
        \midrule
        SIS & $\frac{dx_i(t)}{dt} = -\delta_i x_i(t) + \sum_{j=1}^N a_{ij} x_j(t) (1 - x_i(t))$\\
        \addlinespace
        MP & $\frac{dx_i(t)}{dt} = x_i(t) \left(\alpha_i  - \theta_ix_i(t)\right) + \sum_{j=1}^N a_{ij} \frac{x_i(t)x_j(t)^2}{1+x_j(t)^2}$\\
        
        \addlinespace
        LV & $\frac{dx_i(t)}{dt} = x_i(t) \left(\alpha_i  - \theta_ix_i(t)\right) - \sum_{j=1}^N a_{ij} x_i(t) x_j(t)$ \\
        \addlinespace
        WC & $\frac{dx_i(t)}{dt} = -x_i(t)+\sum_{j=1}^N a_{ij} \frac{1}{1+exp(-\tau(x_j(t)-\mu))}$, \\
        \bottomrule
    \end{tabular}
\end{table}
\begin{itemize}
    \item \textbf{SIS Model: Susceptible–Infected–Susceptible Epidemics}\\
    The SIS model \cite{Kuhl2021}  elucidates the dynamics of epidemics, specifically the spread of infectious diseases within a population. Each nodal state \(x_i(t)\) represents the infection probability of species \(i\). The parameter \(\delta_i\) signifies the curing rate, and the link weight \(a_{ij}\) quantifies the infection rate from species \(j\) to species \(i\). The dynamics of the SIS model capture the intricate interplay between infection and recovery processes, providing a comprehensive understanding of epidemic dynamics and potential control strategies.

    \item \textbf{MP Model: Mutualistic Population Dynamics}\\
    The MP model \cite{harush2017dynamic} explores the intricate dynamics arising from mutualistic interactions within a population of different species. In this context, the nodal state \(x_i(t)\) signifies the population size of species \(i\). The growth parameters \(\alpha_i\) and \(\theta_i\) characterize the intrinsic growth and carrying capacity of species \(i\). Furthermore, the link weight \(a_{ij}\) captures the strength of mutualistic interactions between species \(i\) and \(j\). The dynamics of the MP model reveal how mutualistic relationships influence the growth and interaction patterns among different species, shedding light on the complexity of ecological systems.

    \item \textbf{LV Model: Lotka–Volterra Population Dynamics}\\
    The LV model \cite{macarthur1970species} provides insights into the population dynamics of species engaged in competition. Each nodal state \(x_i(t)\) represents the population size of species \(i\), with growth parameters \(\alpha_i\) and \(\theta_i\) characterizing intrinsic growth and carrying capacity. The link weight \(a_{ij}\) signifies the competition or predation rate of species \(j\) on species \(i\). The dynamics of the LV model showcase the delicate balance between intrinsic growth and the impact of competition or predation, offering a nuanced understanding of species coexistence and competition within ecological systems.

    \item \textbf{WC Model: Wilson–Cowan Neural Firing}\\
    The basic idea behind the Wilson-Cowan model \cite{laurence2019spectral} is to represent the activity of excitatory and inhibitory populations of neurons in terms of their average firing rates. The model assumes that these populations interact with each other through excitatory and inhibitory connections, leading to a dynamic balance between excitation and inhibition.The equations of the Wilson-Cowan model typically take the form of a pair of coupled differential equations, describing the evolution of the firing rates of the excitatory and inhibitory populations over time. The model includes parameters that represent the strengths of the excitatory and inhibitory connections, as well as other factors influencing neural activity.
\end{itemize}

After obtaining the matrix, we apply scPN introduced in Method. Initially, we rank TSP using the previously defined distance metric \( \operatorname{dist}[i, j]=\|x(t_i) - x(t_j)\| + \|\dot{x}(t_i) - \dot{x}(t_j)\| \). Subsequently, we employ the resulting time order to conduct regression analysis and compute the connection matrix \( A \). Iterating these steps yields the correct time sequence \( t_i \) and an alternative connection matrix \( \hat{A} \).

For the piecewise linear dynamic systems, three connection matrices \( A \) of size \( 100 \times 100 \) are randomly generated. Each \( A_{ij} \) denotes the connection between the \( i \)th and \( j \)th nodes in the dynamic network during a specific time period. Subsequently, an initial value \( \mathbf{x}_0 \), a 100-dimensional vector, is randomly generated. The evolving states of 100 nodes over a period of time \( t_1 \) are then computed. The final value of \( t_1 \) serves as the initial value for the subsequent time period \( t_1 \) to \( t_2 \), and similarly, the final value of \( t_2 \) initializes the time period from \( t_2 \) to \( t_3 \). Furthermore, the \( T \) sequences of 100-dimensional vectors are shuffled. Employing the scPN delineated in Method part, both the connection matrices \( A \) and the time order \( t \) can be simultaneously inferred. The experimental outcomes are illustrated in Figure 2(e).

\begin{figure*}[h]
    \centering
    \includegraphics[width=\textwidth]{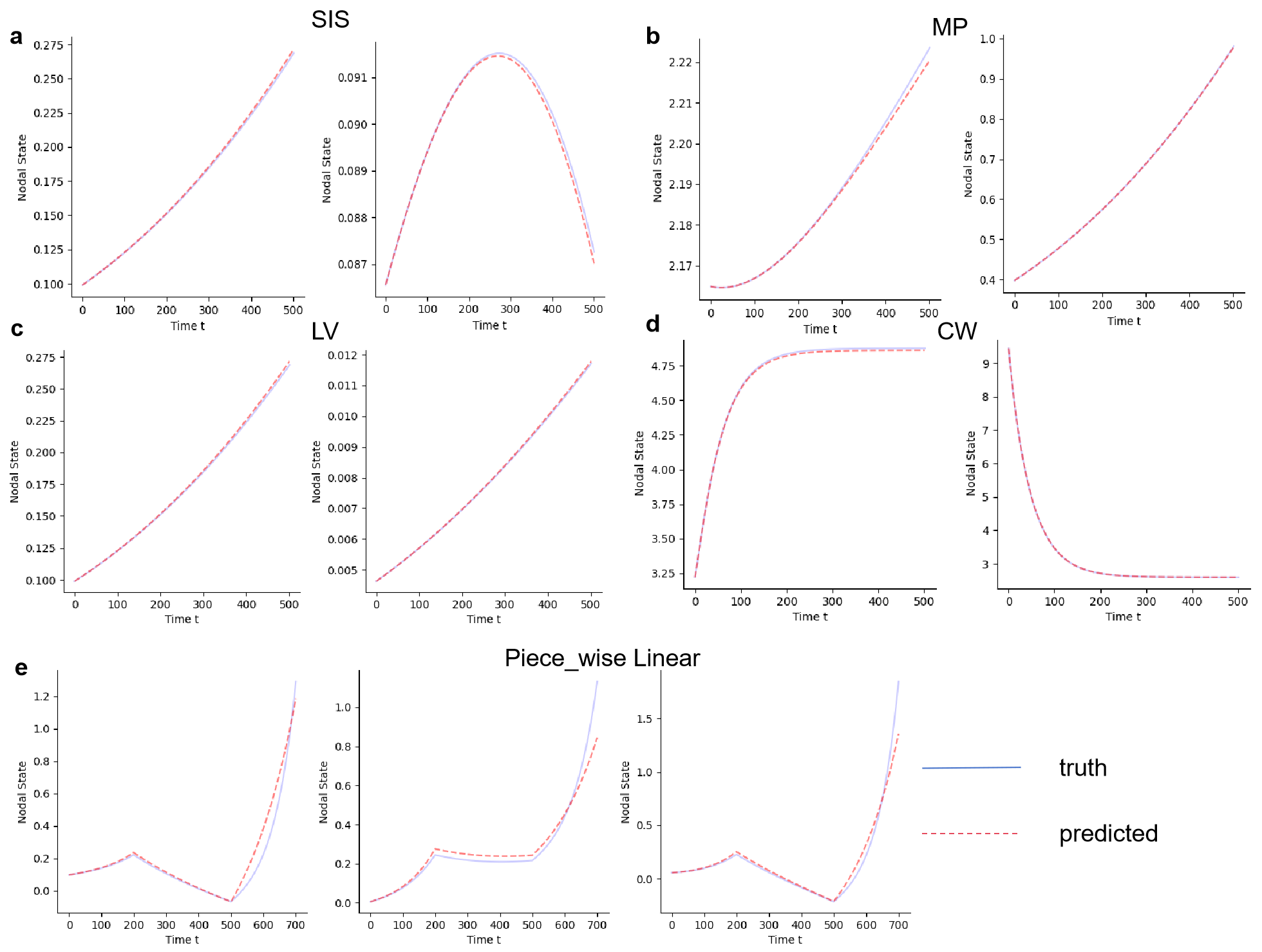}
    \caption{\textbf{The results of scPN on simulation experiments.} Five panels correspond to distinct nonlinear dynamical systems and a piecewise linear system. Within each panel, the blue curve represents true data, while the red curve signifies predicted values. (a$\sim$e) SIS, MP, LV, CW, and piecewise linear network dynamics. To showcase the alignment between model predictions and actual observations, several arbitrary nodes were selected for each  system. The residuals, indicating the deviation between predicted and actual values, are interpolated as $1.9 \times10^{-3}, 1.8\times10^{-2}, 6\times10^{-4}, 3.2\times10^{-2}$, and $3\times10^{-4}$, respectively.}
    \label{fig:1}
\end{figure*}

\subsection*{Pseudotime inference on scRNA-seq datasets with scPN}
In this section, we apply our algorithm to multiple scRNA-seq datasets, focusing primarily on three distinct datasets:
Gastrulation, Oligolite, and Dentate Gyrus Dataset.

Gastrulation is a crucial stage in embryonic development, during which the primary embryo forms three germ layers (endoderm, ectoderm, and mesoderm). During this process, the primary embryo undergoes cell movements and reshaping, leading to the formation of the basic body axes and organ structures necessary for subsequent development.

Dentate gyrus is a part of the hippocampus in the brain, located within the cerebral cortex. It is an area of the brain where neurons are generated, and is closely associated with learning and memory. The main functions of the dentate gyrus include participation in the generation and integration of new neurons, as well as its important role in spatial memory and cognitive functions.

The Oligolite dataset is an open dataset used for studying biological molecules such as DNA, RNA, and proteins. It contains a large amount of molecular sequence data that can be used for bioinformatics and computational biology research.These three dataset is commonly used for training and validating bioinformatics algorithms, such as sequence alignment, genome assembly, and gene expression analysis.

In preprocessing single-cell data for pseudotime ordering, several steps are involved. Initially, quality control measures are applied to filter out low-quality cells and eliminate potential sources of noise. Subsequently, log normalization is performed to mitigate the impact of technical variability and ensure comparability of gene expression levels across cells. High-variance genes are then identified and selected to capture the most informative features of the dataset. Any missing values are imputed to address dropout events or technical noise. Finally, gene expression levels are further normalized to account for variations in sequencing depth and other technical factors. Due to the large size of the dataset, we employ subsampling methods to reduce the number of cells in the entire dataset.

For single-branch datasets, we treat the preprocessed data matrix, with cells as rows and genes as columns, as a shuffled linear dynamic system after preprocessing. Assuming that the time labels of consecutive cells are evenly spaced, our algorithm infers timestamps after shuffling. To address the smoothness of timestamps, we further optimize the time labels using the K-Nearest Neighbors (KNN) method, resulting in the final pseudotime of cells. Detailed results can be observed in Figure 3(d), 3(e).

For multi-branch datasets such as the dentate gyrus dataset, where initial cells diverge into different fates, we cannot treat the entire matrix uniformly. Instead, we first partition the entire dataset using the Leiden algorithm \cite{traag2019louvain}. Subsequently, we address this issue by employing piecewise linear systems, considering each partitioned block of data as a linear dynamic system guided by different connection matrices A. After obtaining pseudotime ordering for the entire dataset, scPN employs the K-nearest neighbors (KNN) method to smooth the results. Detailed results can be observed in Figure 3(f).

\begin{figure*}[h]
    \centering
    \includegraphics[width=\textwidth]{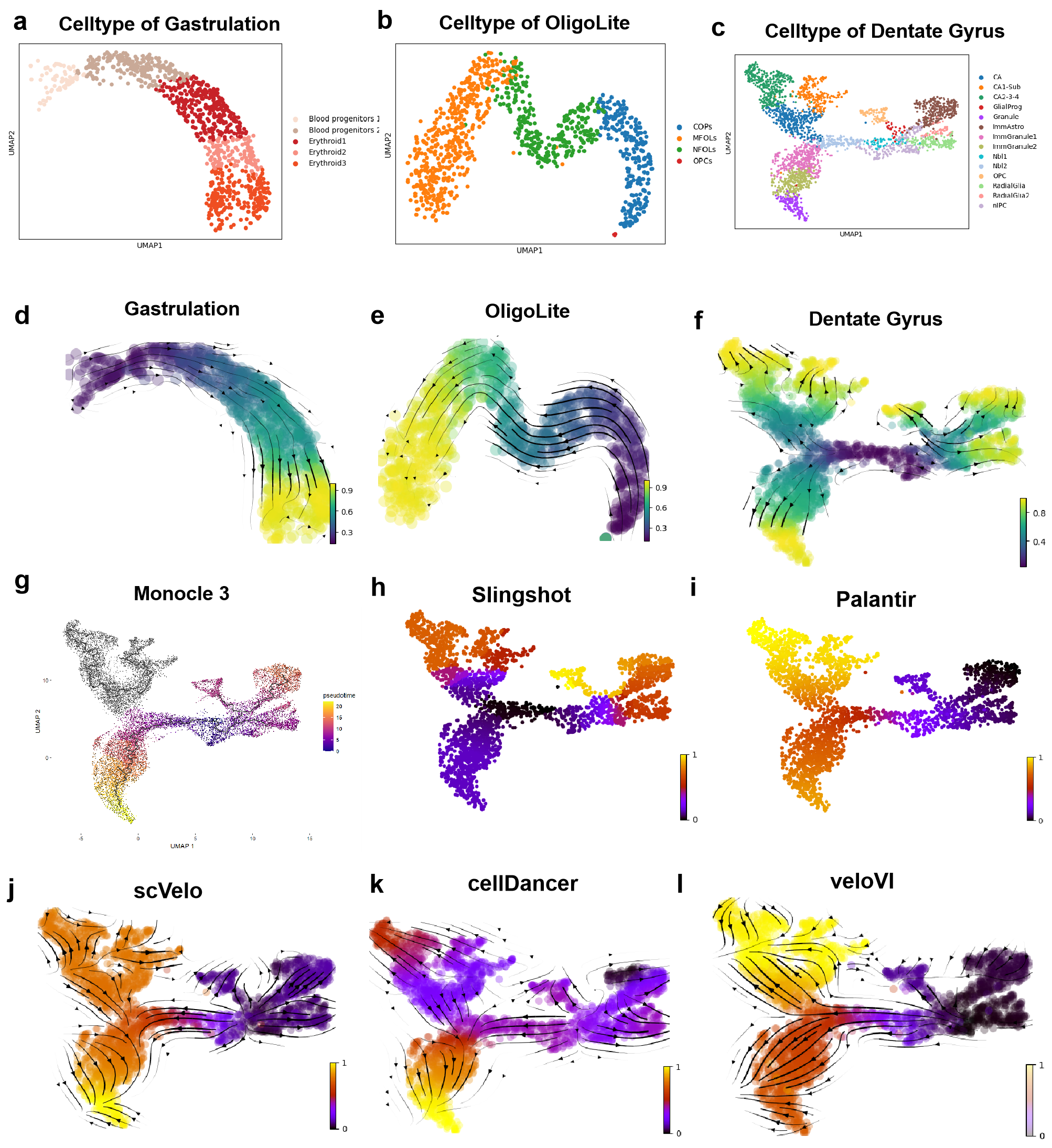}
    \caption{\textbf{The results for real scRNA-seq datasets.} These panels include pseudotime results obtained after applying scPN to three real datasets, as well as the results from three comparison methods.    
    (a$\sim$c) UMAP dimension-reduced cell cluster plots for the Gastrulation, Oligolite, and Dentate Gyrus.Each color represents a different cell type. These plots demonstrate the accuracy of the computed pseudotime.
    (d$\sim$f) scPN obtained pseudotime results and the velocity field for three datasets, named Gastrulation, Oligolite, and Dentate Gyrus. In these three panels, the purple color represents the initial cells, while the yellow color denotes the cells after development.
    The arrows in the figure represent the velocity field, and the results indicate the consistency between pseudotime and velocity.
    (g$\sim$i) The results obtained using Monocle3 (note that the results may differ slightly due to the R implementation, but the overall pseudotime trajectory remains consistent), Slingshot and Palantir.
    (j$\sim$l) scVelo, cellDancer, and veloVI are three methods that calculate pseudotime based on velocity. The figure illustrates the pseudotime and velocity field, with results indicating their inconsistency.  
    }
    \label{fig:3}
\end{figure*}

\subsection*{Comparison with baseline approaches}
In our study, we compare our method with three prominent methods for inferring pseudotime trajectories from single-cell data: Monocle3, Slingshot, and Palantir.

Monocle3 (Figure 3(g)) is a single-cell pseudo-temporal inference method that leverages flow cytometry analysis. It constructs pseudo-temporal trajectories by analyzing changes in cell states and aligns single-cell data onto a common time axis using Dynamic Time Warping (DTW) algorithms. 

Slingshot (Figure 3(h))  is a single-cell pseudo-temporal inference method based on dynamical branching. It infers branching paths of cell development by identifying transition points and branch points in cell states and models cell states using non-linear dimensionality reduction techniques. Slingshot offers an intuitive visualization of branching structures in single-cell data and provides detailed analytical results and interpretations. It excels at identifying branching structures in cell development and can handle complex branching trajectories.

Palantir (Figure 3(i))  is a single-cell pseudo-temporal inference method based on identifying latent branching points. It models the potential trajectories of cell development using transition probabilities between cells and determines major branching points by minimizing changes in these probabilities. Palantir offers an intuitive visualization of single-cell developmental trajectories and excels at identifying branching events, providing detailed analytical results and interpretations.

In the Dentate Gyrus dataset, the inferred cell trajectory and pseudotime  indicates a developmental sequence from Neuroblast 1 (Nbl1) and Neuroblast 2 (Nbl2) to CA2-3-4 neurons, CA1-Sub neurons, Oligodendrocyte Precursor Cells (OPC), Immature Astrocytes (ImmAstro), Granule Cells, and Radial Glia.

This trajectory reflects the differentiation and maturation of cell types within the Dentate Gyrus. It begins with the early neuroblast stages (Nbl1 and Nbl2), progresses through the formation of specific neuronal subtypes (CA2-3-4 and CA1-Sub neurons), and then transitions to OPC and ImmAstro. Subsequently, the trajectory includes the emergence of granule cells, which are among the most abundant neuronal types in the Dentate Gyrus, and Radial Glia, which are precursor cells with multipotency and play crucial roles in embryonic development.

This sequential progression provides insights into the developmental dynamics and cellular composition of the Dentate Gyrus, shedding light on its functional organization and potential roles in neural circuitry and cognition.

Our method successfully inferred the pseudotime trajectories on this dataset, demonstrating significant advantages over three commonly used methods. Slingshot failed to distinguish between CA2-3-4 and CA1-Sub cell types accurately, while Monocle3 was unable to include CA2-3-4 and CA1-Sub data in its principal graph plots. Additionally, Palantir and Slingshot defaulted to assuming cell differentiation from one segment to another based solely on UMAP plots. These limitations highlight the inability of existing methods to accurately capture the pseudotime dynamics of single-cell trajectories with diverse cell fate.
\subsection*{Consistency between pseudotime and velocity}
The scPN derives the single-cell velocity from the left-hand side of Equation (2), \(\frac{d x_i(t)}{d t}\). This enables the visualization of the velocity field on the pseudotime plot (Figure 3(d) (e) (f)).Based on pseudotime, scPN calculates the temporal loss for all genes and selects the top 500 genes. After obtaining a velocity matrix of size \(500 \times n\) (where \(n\) is the total number of genes), we use scVelo to visualize the velocity field on a UMAP plot.Finally, we removed some velocities with smaller norms, setting the parameter min\_mass = 3.0 for this purpose.

scVelo is a popular tool for analyzing RNA velocity in single-cell transcriptomics. It extends the concept of RNA velocity by incorporating dynamical modeling to recover cell-specific velocities and infer pseudotime. The method uses velocity estimates to reconstruct cell lineage trajectories by projecting velocities onto low-dimensional embeddings such as UMAP.

cellDancer is another method designed to infer cellular dynamics and pseudotime from velocity data. By estimating the RNA velocity of individual cells, it predicts the future state of cells and infers the developmental trajectory based on these velocity vectors. Similar to scVelo, cellDancer relies on velocity to determine the direction of cell transitions over time.

veloVI is a more recent method that builds upon the idea of velocity-based pseudotime inference, optimizing the reconstruction of cellular dynamics. It leverages RNA velocity to identify branching points in the developmental process and infer pseudotime in multi-lineage differentiation scenarios.Since veloVI does not have a built-in method for pseudotime inference, we first used veloVI to obtain the velocity estimates and then employed scVelo to generate the pseudotime (Figure 3(l)).

All three methods scVelo, cellDancer, and veloVI derive pseudotime from velocity by first calculating the velocity field and then inferring the pseufotime of cells based on this field. However, this sequential approach can lead to inconsistencies between the inferred pseudotime and the velocity field (Figure 3(j) (k) (l)).

\begin{figure*}[h]
    \centering
    \includegraphics[width=\textwidth,height=15cm]{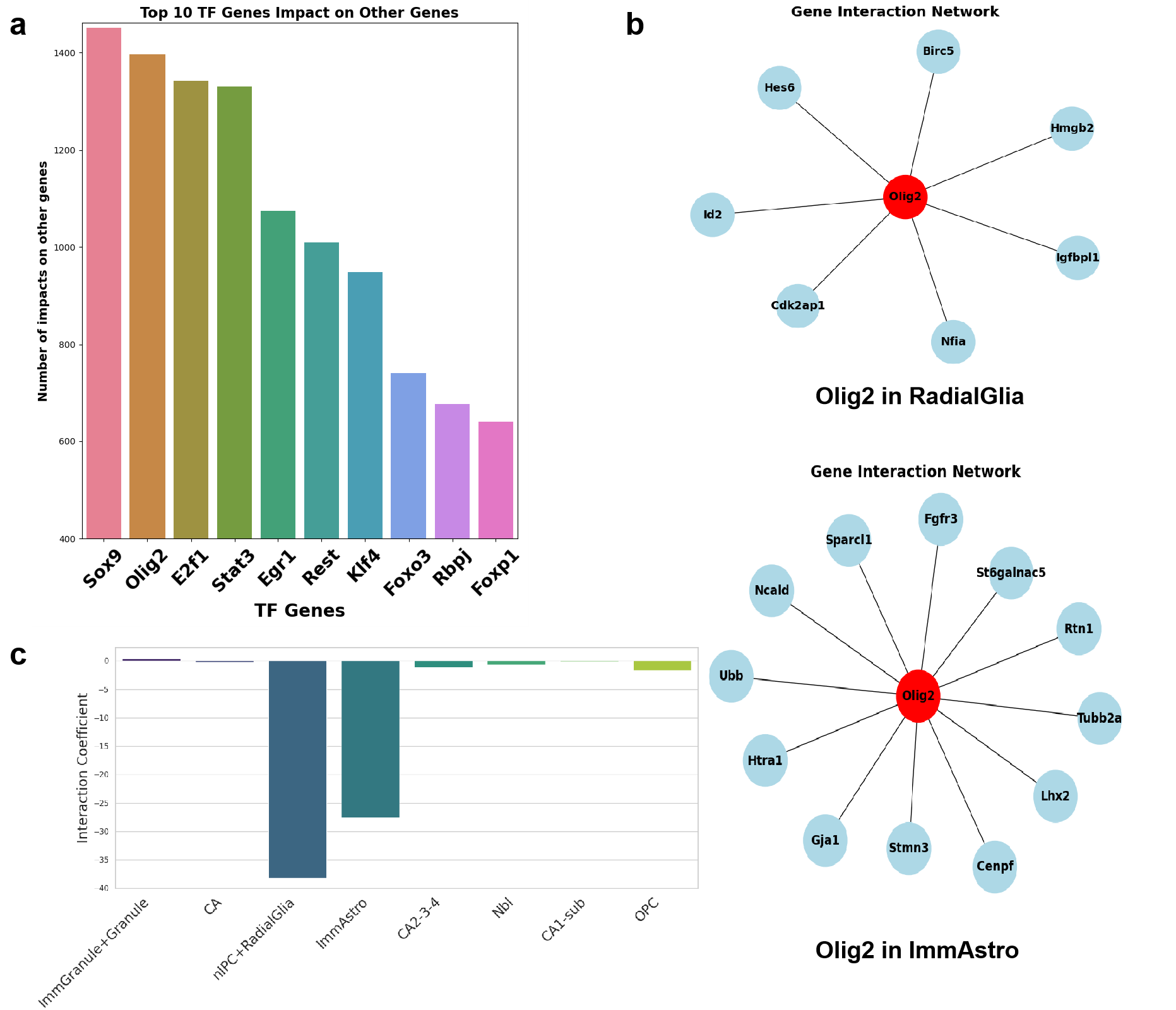}
    \caption{\textbf{The Explanation of learned gene interaction matrix.} (a) presents a bar plot of the number of TF genes influencing other genes across all clusters. scPN removes edges with smaller weights from the gene connectivity matrix and then calculated the degree and then finds several important TF genes such as Sox9, Olig2, Stat3, and Klf4 can be identified. (b) The impact of the TF gene Olig on other genes in Radial Glia and ImmAstro.  (c) The interaction coefficients between Klf4 and Stat3 across different cell types.}
    \label{fig:4}
\end{figure*}

\subsection*{Key Regulatory Identification from Gene Interaction Matrix Learned by scPN}
Based on these cluster-specific gene interaction matrices learned by scPN, we can identify the key TFs by the following approach. we first normalized the gene interaction matrix calculated by scPN. Then, we identified the maximum value and set connection values smaller than 85\% of the maximum value to zero, meaning we only focus on edges with larger connection coefficients. Finally, we ranked the nodes by the sum of degrees in all matrices and selected the top nodes.

On the Dentate Gyrus dataset, the top-10 key TFs identified by scPN are Sox9 \cite{xu2015radial}, Olig2 \cite{ohyama2023psmad3+}, E2f1, Stat3 \cite{chen2017homocysteine}, Egr1, Rest \cite{nassar2023repressor}, Klf4 \cite{park2014induction}, Foxo3, Rbpj, Foxp1 (Figure 4(a) ), most of which have been verified to be significant for neural system development. Especially, Sox9, which as a high-impact transcription factor present across multiple clusters, has been verified as a significant TF across various cell clusters, including Radial Glia \cite{xu2015radial}, Neuroblasts (NBl), Astrocytes \cite{sun2017sox9}, and Oligodendrocyte Progenitor Cells (OPC) \cite{ng2022sox9}.

Sox9 is particularly crucial in Radial Glia and Astrocytes, where it regulates the transition from neurogenesis to gliogenesis. Previous studies have highlighted that Sox9 expression is vital in Radial Glia for their role in neurogenesis and gliogenesis, as these cells transition to Astrocytes and other glial cell types \cite{vong2015sox9}. Sox9 is essential in maintaining the stem cell properties of Radial Glia and their ability to differentiate into neurons and Astrocytes as needed. Furthermore, Sox9's role extends to its involvement in the maturation and differentiation processes within the neuroblast and OPC populations, ensuring the proper formation and functioning of these cells within the dentate gyrus. This regulatory function of Sox9 ensures the proper development and maintenance of neural and glial cells, contributing to the overall structural and functional integrity of the hippocampus.

scPN can also be utilized for analyzing differences in TF activity across various cell types. For example, scPN identifies Olig2 as the second most significant TF. Furthermore, scPN suggests that its activity varies between two distinct clusters: Astrocytes and Radial Glia.  As shown in Figure 4(b), Olig2 emerged as a critical TF gene for Astrocytes, but not for Radial Glia. This differential expression suggests that Olig2 serves as a reliable marker to distinguish Astrocytes from Radial Glia. Our findings are corroborated by a previous study \cite{ohyama2023psmad3+} which emphasizes that Olig2 expression distinctly marks Astrocytes, differentiating them from Radial Glia. This distinction is crucial for understanding the cellular composition and lineage relationships within the hippocampal dentate gyrus.

Klf4, another top-ranked key TF suggested by scPN, also plays a significant role as a TF in the dentate gyrus, especially in the context of Astroglial response to ischemic injury. Several studies \cite{park2014induction}  highlight its involvement in the regulation of Astrocyte activation and neuroinflammatory processes following such injuries.For instance, Klf4 has been shown to regulate the activation of reactive Astrocytes, which are critical for brain response after ischemic stroke. This regulation is essential for managing neuroinflammation and promoting recovery processes in the
brain \cite{wang2023critical} \cite{yin2015krupple}. Furthermore, Klf4's expression is influenced by cerebral ischemia, where it helps to modulate inflammation and maintain the integrity of the blood-brain barrier through the regulation of tight junction proteins \cite{zhang2020Klf4}.

Furthermore, in hippocampal neurons, Klf4 acts as a negative regulator of Stat3 activity, shown in Figure 4(c). Studies \cite{sahin2020leptin} show that leptin-induced synaptogenesis involves Klf4 upregulation, which leads to the inhibition of Stat3 signaling. Our results align with this by revealing significant negative coefficients between Stat3 and Klf4, particularly in Nbl and Radial Glia, further supporting their antagonistic relationship.

\section*{Discussion}
In this study, we proposed a novel algorithm scPN for single-cell analysis, focusing on generating interpretable, piecewise ODE models of gene regulatory network dynamics. Our method presents several advantages. Firstly, it is capable of computing both the temporal sequence and the gene regulatory network simultaneously, providing a comprehensive view of the dynamics in cellular development. Secondly, it effectively addresses the challenges posed by branching single-cell trajectories, where cells may differentiate into different lineages. This feature enhances the accuracy and interpretability of the models in representing complex cellular differentiation processes.

The interpretability of our results is a key advantage of our approach. Unlike deep neural network algorithms, which often function as "black boxes", our method scPN employs basic optimization algorithms, making the resulting models more interpretable. Additionally, by incorporating knowledge about TF and target genes, our algorithm can identify key TF genes based on the dataset. This integration of domain knowledge not only improves the interpretability of the models but also ensures that the identified regulatory relationships are biologically meaningful.

Our scPN is theoretically supported by the fact that each step is designed to reduce the loss function, ensuring convergence to a local minimum. This theoretical foundation guarantees the reliability and robustness of our method. Furthermore, we conducted simulation experiments to demonstrate the feasibility and effectiveness of our approach. These simulations confirmed that our algorithm can accurately model gene regulatory dynamics and handle the complexity of single-cell trajectories.

In comparison with related work, we evaluated scPN against popular single-cell temporal analysis methods such as Monocle3, Palantir, and Slingshot. On the multi-branching dataset Dentate Gyrus, our method produced more accurate results and also generated a gene-gene correlation matrix. This indicates that our approach has significant advantages in handling complex single-cell datasets.Additionally, we will compare scPN with well-known single-cell velocity methods: scVelo, cellDancer, and veloVI. Since scPN employs an EM approach to simultaneously optimize both pseudotime and velocity, it ensures consistency between the two. In contrast, scVelo, cellDancer, and veloVI derive pseudotime from velocity, which can lead to inconsistencies between the pseudotime and the velocity field.

In terms of limitations, scPN is similar to the EM algorithm and requires a good initial temporal sequence. Additionally, when calculating the temporal sequence with a known connection matrix, we used a heuristic algorithm. Therefore, the computation time can be particularly long when dealing with a very large number of cells.

Overall, scPN provides a robust and interpretable framework for analyzing gene regulatory network dynamics from scRNA-seq data, addressing the branching challenges. Future work will focus on further enhancing the algorithm's adaptability and exploring its application in various biological contexts.
\section*{Data and Code Availability}

The gastrulation erythroid dataset \cite{pijuan2019single}, derived from the transcriptional profiles of mouse embryos, encompasses the expression levels of 53,801 genes across 9,815 cells. This dataset is accessible via the scVelo.datasets.gastrulation\_erythroid() function in the scVelo package. Additionally, the 10x Genomics embryonic mouse brain dataset can be found at the 10x Genomics website:
\href{https://www.10xgenomics.com/resources/datasets/fresh-embryonic-e-18-mouse-brain-5-k-1-standard-1-0-0}{Gastrulation Dataset}.
The oligodendrocyte differentiation dataset was obtained from a survey of the mouse nervous system (Sequence Read Archive (SRA) accession SRP135960).
The data from mouse P0 and P5 hippocampus was extracted from dataset \cite{hochgerner2018conserved}, available from the Gene Expression Omnibus (GEO) under accession GSE104323. The loom data is available at:
\href{http://pklab.med.harvard.edu/velocyto/DentateGyrus/DentateGyrus.loom}{Dentate Gyrus Dataset}.
The ChEA TF-target database is available at \cite{lachmann2010chea}:
\href{https://maayanlab.cloud/Harmonizome/dataset/CHEA+Transcription+Factor+Targets}{ChEA TF-target database}.
The code to reproduce results, together with documentation and examples of usage, is available on GitHub at \href{https://github.com/ZHOUZHEN2002/scPN/tree/master}{https://github.com/ZHOUZHEN2002/scPN/tree/master}

\acknow{ 
This work was supported by the National Natural Science Foundation of China (No. 62073219, 92059206), and the Science and Technology Commission of Shanghai Municipality (22511104100).}

\showacknow{} % Display the acknowledgments section

% \bibsplit[2]
%Use \bibsplit to split the references from the body of the text. Value "[2]" represents the number of reference in the left column (Note: Please avoid single column figures & tables on this page.)

% Bibliography
\bibliography{pnas-sample}

\end{document}